\documentclass[sn-basic]{sn-jnl}
\usepackage[caption=false,font=footnotesize]{subfig}
\usepackage{caption}
\usepackage{float}
\usepackage{array}
\usepackage{tabularx}
\usepackage{booktabs}

\jyear{2021}%
\theoremstyle{thmstyleone}%
\theoremstyle{thmstyletwo}%

\theoremstyle{thmstylethree}%

\begin{document}

\title[Automatic Endoscopic Ultrasound Station Recognition with Limited Data]{Automatic Endoscopic Ultrasound Station Recognition with Limited Data}

\author[1]{\fnm{Abhijit} \sur{Ramesh}}

\author[1]{\fnm{K N} \sur{Anantha Nandanan}}
\equalcont{These authors contributed equally to this work.}

\author[1]{\fnm{Nikhil} \sur{Boggavarapu}}
\equalcont{These authors contributed equally to this work.}

\author[2]{\fnm{Priya} \sur{Nair}}

\author[1]{\fnm{Gilad} \sur{Gressel}}\email{gilad.gressel@am.amrita.edu}

\affil[1]{\orgdiv{Amrita Center for Cybersecurity Systems \& Networks}, \orgname{Amrita Vishwa Vidyapeetham}, \orgaddress{\street{Amritapuri}, \country{India}}}

\affil[2]{\orgdiv{Department of Gastroenterology}, \orgname{Amrita Institute of Medical Sciences}, \orgaddress{\city{Kochi}, \country{India}}}


\abstract{Pancreatic cancer is a lethal form of cancer that significantly contributes to cancer-related deaths worldwide. Early detection is essential to improve patient prognosis and survival rates. Despite advances in medical imaging techniques, pancreatic cancer remains a challenging disease to detect. Endoscopic ultrasound (EUS) is the most effective diagnostic tool for detecting pancreatic cancer. However, it requires expert interpretation of complex ultrasound images to complete a reliable patient scan. To obtain complete imaging of the pancreas, practitioners must learn to guide the endoscope into multiple ``EUS stations" (anatomical locations), which provide different views of the pancreas. This is a difficult skill to learn, involving over 225 proctored procedures with the support of an experienced doctor. We build an AI-assisted tool that utilizes deep learning techniques to identify these stations of the stomach in real time during EUS procedures. This computer-assisted diagnostic (CAD) will help train doctors more efficiently. Historically, the challenge faced in developing such a tool has been the amount of retrospective labeling required by trained clinicians. To solve this, we developed an open-source user-friendly labeling web app that streamlines the process of annotating stations \emph{during} the EUS procedure with minimal effort from the clinicians. Our research shows that employing only 43 procedures with no hyperparameter fine-tuning obtained a balanced accuracy of 89\%, comparable to the current state of the art. In addition, we employ Grad-CAM, a visualization technology that provides clinicians with interpretable and explainable visualizations.

}


\maketitle

\section{Introduction}\label{sec1}
Pancreatic cancer is the seventh leading cause of cancer-related deaths worldwide~\cite{Bray-F-cancer-stats}. Early detection is essential to improve the prognosis and survival rate of patients with pancreatic cancer. Despite advances in imaging technology, the survival rate remains low, with a reported 12\% survival rate~\cite{Rawla-P-Epidemiology}. This is because pancreatic cancer is often asymptomatic until it reaches an advanced stage, making early detection unlikely. 

Magnetic Resonance Imaging (MRI), Computed Tomography (CT) scans, Endoscopic Ultrasound (EUS), and Positron Emission Tomography (PET) scans are different medical imaging techniques used for diagnosing pancreatic cancer. Among these imaging techniques, EUS uses high-frequency ultrasound waves to produce detailed images of the pancreas and surrounding organs. EUS is considered the most effective method for detecting early pancreatic cancer because it provides the most accurate visual of the size, location, and extent of the tumour~\cite{Gonzalo-Marin-Role-endoscopic}. A significant advantage of EUS is its capacity to detect very small tumors, as small as 2mm - 3mm in size; in comparison, CT and MRI can only detect tumors larger than 1cm. The EUS procedure demands a high level of expertise and experience, involving over 225 proctored procedures to be done before being assessed for competency~\cite{faulx2017guidelines}. In other forms of US imaging, the probe location is fixed and easy to control because it is in the examiners hand. However, during EUS, clinicians trained in endoscopy need to interpret complex ultrasound images in real-time using a flexible probe in a constantly moving environment. To obtain complete imaging of the pancreas, practitioners must learn to guide the endoscope into multiple ``EUS stations" (anatomical locations), which provide different views of the pancreas. The recognition of the EUS stations is crucial to the EUS procedure as it enables targeted biopsies, accurate diagnosis, and aids in further surgical planning and monitoring.

In order to assist doctors in learning the EUS procedure, previous studies have demonstrated the feasibility of computer-aided diagnostic (CAD) systems that use deep learning techniques in order to identify the pancreatic stations and whether or not the tumor is cancerous~\cite{automatic-video, CAD-2}. However, these studies required retrospective annotated data from expert clinicians, increasing the clinician's workload. To solve this, we have developed and open-sourced a labeling application that streamlines the process for the Endoscopist to annotate the pancreas station during the EUS procedure, adding nearly no effort to the clinician. This type of ``real-time" labeling is successful because we train our CAD system on all the data found in the video. We do not require the Endoscopist to only select gold-standard images and manually remove difficult images.

We build an explainable AI-assisted tool to help Endoscopists identify different stations during the procedure. Incorporating this AI-assisted tool can improve the accuracy of diagnoses and decision-making in treating pancreatic cancer. Importantly, we demonstrate that a state-of-the-art system can be built with limited data and little labeling effort from the clinicians. By reducing the data requirements for training models, we aim to democratize these complex CAD systems. Our study utilized only 43 EUS procedures, accounting for approximately 10-15\% of the data utilized in other related studies~\cite{Zhang-Jun-video, Zhang-J-segmentation}. We leverage preprocessing techniques on the EUS images to enhance image quality, thereby improving the overall performance of the model~\cite{precise-image-preprocessing}. We also incorporate Grad-CAM visualizations to provide insight into the decision-making process of deep learning models, thus producing an explainable CAD system~\cite{Grad-CAM, gradcam-for-medical}. This also enables the CAD system to be used as an offline training program, where doctors can practice identifying stations retrospectively. Our experiments show that using preprocessing techniques, we achieve an accuracy of 89.0\%, and without using the preprocessing techniques, we achieve an accuracy of 87.6\%.\\

We make the following novel contributions:
\begin{itemize}
    \item Achieve a balanced accuracy of 89\% with only preprocessing the input and zero fine-tuning on a small dataset; this is comparable to state-of-the-art that uses transfer learning, fine-tuning, and larger datasets~\cite{bile-duct, Zhang-Jun-video, Zhang-J-segmentation}.
    \item Utilize Grad-CAM to provide explainability to the physicians during procedures.
    \item Develop and open-source an EUS labeling application that allows clinicians to annotate station and FNA timestamps during pancreas ultrasound procedures.~\href{https://github.com/Amrita-Medical-AI/eusml-labeller}{Github Repository}.
    
\end{itemize}

\section{Related Work}\label{sec2}

In recent years, deep learning algorithms have shown great promise in detecting various diseases from medical imaging data, including ultrasound images~\cite{bile-duct}. Studies have shown that deep learning models perform comparably to or better than human healthcare professionals (HCPs) in most cases, particularly in the detection of diseases such as skin cancer, breast cancer, lung cancer, and diabetic retinopathy. These findings suggest that deep learning algorithms may be effective in the task of pancreas station classification using ultrasound images~\cite{comparison-of-deep-learning}. 

Yao et al.~\cite{bile-duct}, proposed a framework consisting of a bile duct(BD) segmentation network and station recognition network to classify biliary tract diseases in EUS images. The framework achieved a classification accuracy of 93.3\% for BD stations and an F1 score of 0.77 for segmentation for the internal validation set. For classification, the model attained an accuracy of 82.6\% for the external validation set. This highlights the potential for deep learning in the context of pancreatic cancer diagnosis and staging. 

Several other studies have investigated deep-learning techniques for pancreatic station classification. Zhang et al.~\cite{Zhang-Jun-video} proposed a transfer learning approach using a pre-trained ResNet model to classify six stations of the pancreas. The authors achieved an accuracy of 82.4\% on the external dataset and an accuracy of 86.2\% on the per-frame video test. The analysis involved selecting the best images from 311 videos. In our study with 43 videos, less than 15\% of their dataset, we achieved an accuracy of 90\% on the test dataset. 

LU et al.~\cite{Zhang-J-segmentation} proposed a two-stage framework where a convolutional neural network (CNN) was first used to detect the pancreas in the EUS images. Then, a region-based CNN was trained to classify the pancreas into different EUS stations. The authors achieved an accuracy of 95.6\% in station recognition. To conduct their investigation, the authors utilized a dataset consisting of 269 procedures, roughly six times our dataset.
 
Building on this, Jarmillo et al.~\cite{María-Jaramillo} proposed a novel approach for automatically detecting pancreatic tumors in endoscopic ultrasound (EUS) videos using transfer learning. The authors used pre-trained CNN models to classify cancerous pancreatic tumors vs. benign pancreatic tumors and achieved an accuracy of 93.2. The CNN was trained on a dataset of 66,249 EUS images from 55 videos and was evaluated on a test set of 22,097 images from 18 videos. They used pre-processing techniques to remove noise and enhance the tumor region, resulting in improved accuracy and reduced variability in image quality. 

None of the prior studies systematically experimented with different preprocessing techniques in order to enhance the performance of deep learning models on EUS datasets. All previous work relied on large datasets retrospectively annotated by human clinicians, making scalability difficult. 

Despite working with a smaller dataset, comprising just approximately 10-15\% of the videos used in comparative studies, we achieved a balanced accuracy of 89\% for the test set. Our data labeling process did not involve manually annotating ``gold standard" images, making it less time-consuming compared to previous approaches. We also introduce the Grad-CAM visualization technique to provide clinicians with transparent and explainable results. This allows for a better understanding of the model's decision-making process and facilitates trust for medical professionals.

\begin{table}
\centering
\caption{Comparison of EUS Datasets for Station Classification}
\resizebox{\columnwidth}{!}{%
\begin{tabular}{@{}crrrl@{}}
\toprule
\multicolumn{1}{c}{Paper} & 
\multicolumn{1}{l}{N. Patients} & 
\multicolumn{1}{l}{N. Images} & 
\multicolumn{1}{l}{Performance} & 
\multicolumn{1}{l}{Hyperparameter Tuning} \\ 
\cmidrule(l){1-5} 
\cite{Zhang-J-segmentation} & 269 & 18,061 & 94.1\% & Not Specified \\ 
\cite{Zhang-Jun-video} & 311 & 19,486 & 94.2\% & Fine-tuning \\ 
\cite{María-Jaramillo} & 55 & 66,249 & 93.2\% & Grid search \\ 
\cite{automatic-video} & 41 & 179,092 & 66.8\% & Not Specified \\ 
Our Work & 43 & 16,081 & 90\% & None \\ 
\bottomrule
\end{tabular}%
}
\end{table}

\subsection{Preprocessing methods}\label{prepocessing}

\begin{figure}
  \centering
  \subfloat[CLAHE]{\label{fig:CLAHE-EUS}\includegraphics[width=0.30\textwidth]{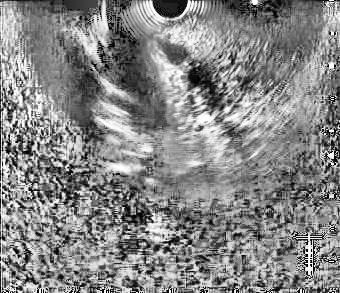}}
  \hfill
  \subfloat[Quantile Capping]{\label{fig:Quantil-EUS}\includegraphics[width=0.30\textwidth]{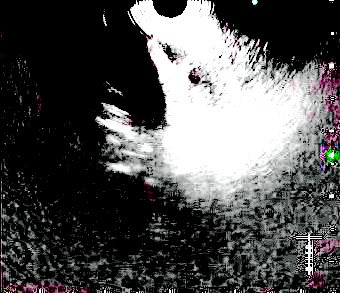}}
  \hfill
  \subfloat[Gaussian Smoothing]{\label{fig:Guassian-EUS}\includegraphics[width=0.30\textwidth]{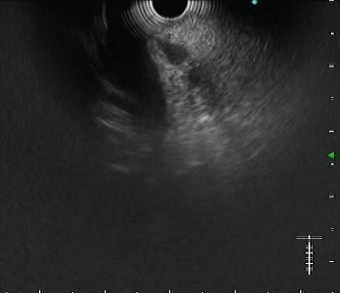}}
   \vskip\baselineskip
\subfloat[Fourier Transform]{\label{fig:Fourier-EUS}\includegraphics[width=0.30\textwidth]{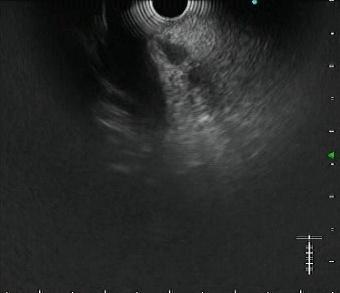}}
\hspace{1cm}
\subfloat[Denoising]{\label{fig:Denoising-EUS}\includegraphics[width=0.30\textwidth]{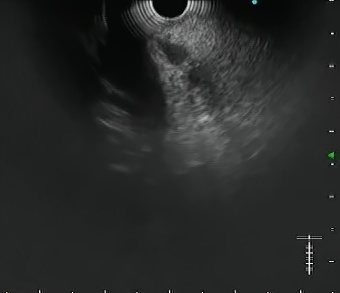}}
  \captionsetup{justification=centering}
  \caption{Transformed EUS Image under different Preprocessing techniques}
  \label{fig:both2}
\end{figure}

As with all machine learning, enhancing data quality will result in improved performance. Due to the nature of video recording, artifacts, and additional noise are often captured in a recording, which degrades the quality of the video. We employ image preprocessing techniques to increase contrast, remove noise, and smoothen/blur the image.

\begin{itemize}
    \item \textbf{Contrast-Limited Adaptive Histogram Equalization:} Contrast-limited adaptive histogram equalization (CLAHE) is an image enhancement technique that is widely used to improve the contrast and brightness of images~\cite{fuzzy-clahe}. It is a variant of adaptive histogram equalization (AHE), a non-linear contrast enhancement method. The idea behind histogram equalization is to transform an image's intensity histogram such that the distribution of the pixel intensities is spread out more evenly across the intensity range, thereby enhancing the contrast of the image. However, histogram equalization may produce undesirable results, such as noise amplification in regions of low contrast. CLAHE was introduced to address this limitation by dividing the image into smaller regions and equalizing the histogram separately for each region. This results in a more localized contrast enhancement, which can help preserve the details of the image~\cite{Pizer-clahe}.

    \item \textbf{Gaussian Smoothing:} Image filtering techniques, such as Gaussian smoothing, are commonly used in image denoising to reduce high-frequency noise while preserving image edges and structures. Gaussian smoothing involves convolving an image with a Gaussian kernel, a bell-shaped function that assigns weights to neighbour pixels based on their distance from the center pixel. The smoothing effect of the Gaussian kernel is determined by its standard deviation, with a higher standard deviation resulting in a more significant blur. Xiao et al.~\cite{Xiao-gaussian} presented a study on developing an automatic brain MRI segmentation scheme with Gaussian smoothing. The paper highlighted the significance of Gaussian smoothing in medical imaging, showcasing its ability to preprocess images, effectively reducing noise, and enhancing the clarity of the features in the image.

    \item \textbf{Quantile Capping:} Quantile capping is a technique used in image processing to limit the dynamic range of an image by capping the extreme pixel values. This is done by finding the upper and lower quantiles of the pixel intensity distribution and capping the values outside this range. This process can help improve the visual quality of an image and enhance its contrast. This technique converts the features into a normal distribution, spreading out the most common values for a particular feature. It also lowers the impact of marginal outliers.

    \item \textbf{Denoising:} In image processing, denoising techniques are used to remove unwanted noise from images. Non-local Means-Based Denoising(NLM) algorithm is one such image denoising algorithm that utilizes the self-similarity of images to remove noises, while still preserving the important image artifacts. it works by comparing each pixel in the image to all other pixels, and then the average of these similar pixels is used to get the effective denoised pixel value. Heo et al. \cite{heo2020image} undertook a systematic review to determine the effectiveness on using NLM algorithm for denoising (Magnetic Resonance)MR images. The study shows that not only was it effective at removing noises from the MR images while still keeping the important image artifacts, but also outperformed other denoising algorithms in terms of peak signal-to-noise ratio(PSNR) value, which is quality meansurement in which higher the PSNR value, better the image quality.  

    \item \textbf{Fourier Transform:} The Fourier Transform is a mathematical tool for decomposing a signal into its frequency components. In image processing, the Fourier Transform converts an image from its pixel-based spatial domain to its frequency-based domain. This transformation helps analyze the image in terms of its frequency content. Working in the frequency domain has several advantages, such as filtering or enhancing specific frequencies, which can be performed more efficiently. Fourier Transform is used in image compression and for the removal of noise in an image. To remove noise in an image, a mask is applied to the transformed image to suppress the noise frequency components while preserving the desired image information. Various strategies can be employed to design effective masks for noise removal in images. This includes utilizing masks of different shapes and sizes, targeting specific regions of the transformed image, or applying thresholding techniques to identify and suppress noise-containing frequencies. After applying a filter to remove noise frequency components, the inverse Fourier Transform transforms the modified frequency-based image into the spatial domain.

\end{itemize}

\subsection{CNN}
The convolutional neural network (CNN)~\cite{deeplearning-review} is widely used and considered the most effective for medical image classification tasks~\cite{CNN}. The CNN is a powerful feature extractor; therefore, it can be used to classify medical images and avoid complex and expensive feature engineering. 
 
We used three classical convolutional neural network architectures: ResNet, DenseNet, and EfficientNet.
A description of these architectures is presented below.

\begin{itemize}
    \item \textbf{ResNet:} This architecture solves a common problem in deep learning called the vanishing gradient problem. This problem occurs when deep neural networks have many layers, which makes it difficult for them to learn effectively. ResNets addresses this problem by using a technique called skip connections. The skip connection connects non-contiguous layers using a direct connection. These connections act like shortcuts, allowing information to flow easily through the network. By adding these shortcuts, the network can train deeper and increase performance on classification tasks. 
 
    The number of trainable parameters of ResNet18 and ResNet34 is 11M and 63.5M, respectively.
    \item \textbf{DenseNet:} This architecture uses dense connections between layers through Dense Blocks, which directly connect all layers (with matching feature-map sizes) to one another. Each layer obtains additional inputs from all preceding layers and passes its feature maps on to all subsequent layers. These dense connections help the model gain collective knowledge with fewer layers (with fewer parameters), which helps the model learn faster. 

    The number of trainable parameters of DenseNet121, DenseNet161 and DenseNet201 is 8M, 29M, and 20M, respectively.
    \item \textbf{EfficientNet:} This architecture manages the tradeoff between accuracy and computational cost. It achieves this by using a compound scaling technique that scales the network's depth, width, and resolution in a principled manner. Unlike conventional practice that uses arbitrary values to scale these factors. The EfficientNet scaling method uniformly scales network width, depth, and resolution with a set of fixed scaling coefficients $\alpha$, $\beta$, and $\gamma$. These scaling coefficients are determined by a small grid search on the original small model~\cite{EfficientNet}. The trainable parameters of EfficientNetB0 and EfficientNetB3 are 5.3M and 12M, respectively.
\end{itemize}

\section{Methodology}\label{sec4}
\subsection{Dataset}\label{dataset}
 
\begin{figure}
  \centering
  \subfloat[Label Tool Screen 1]{\includegraphics[width=0.3\textwidth]{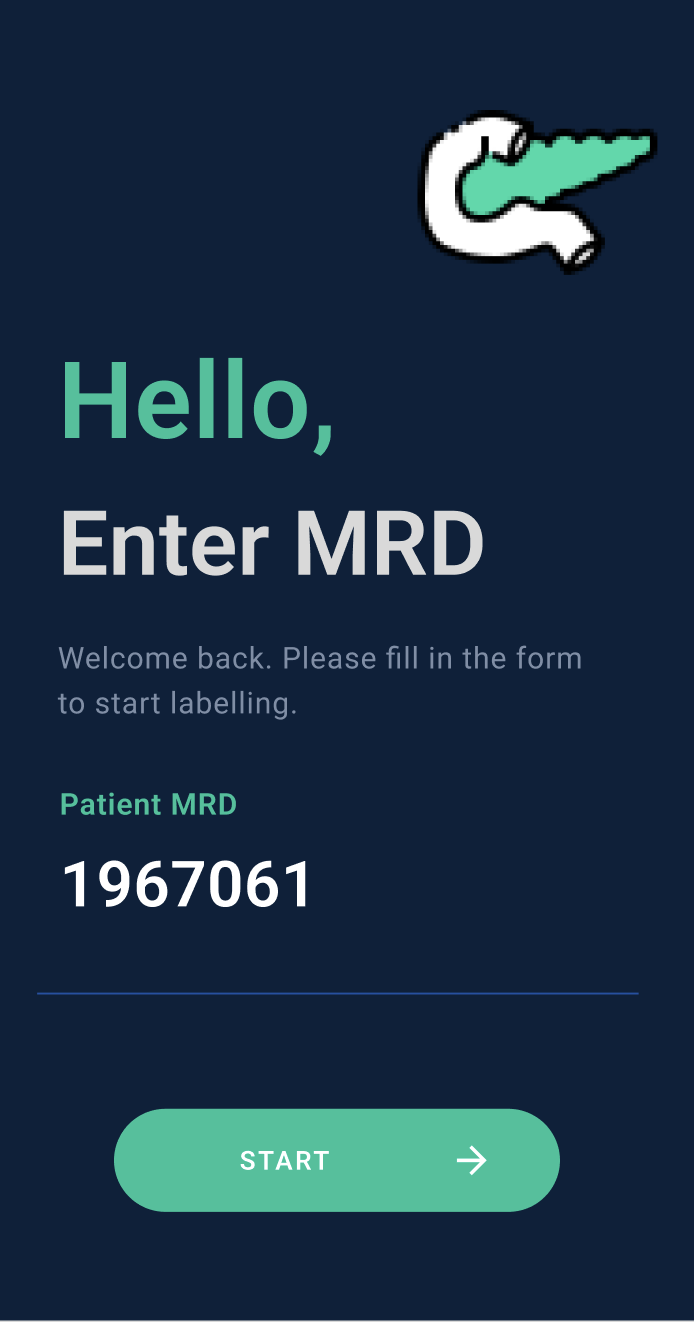}\label{fig:Label-Tool-Screen-1}}
  \hfill  
  \subfloat[Label Tool Screen 2]{\includegraphics[width=0.3\textwidth]{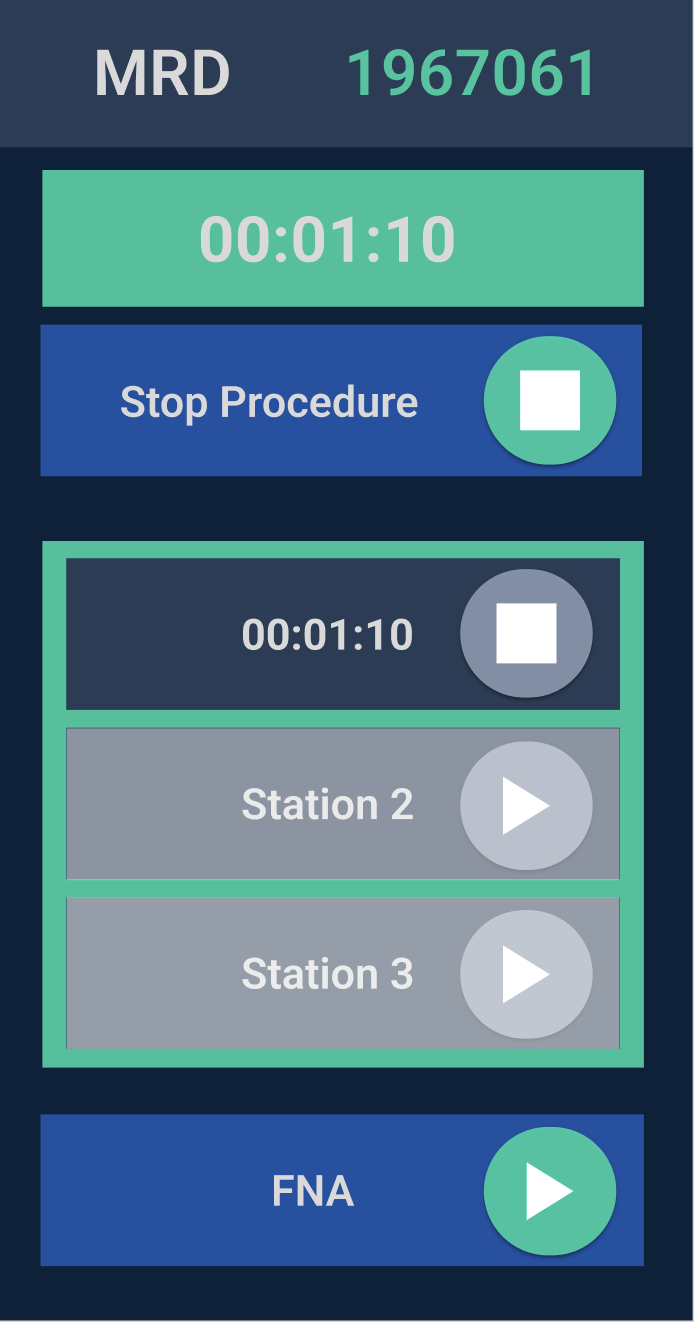}\label{fig:Label-Tool-Screen-2}}
  \hfill
  \subfloat[Label Tool Screen 3]{\includegraphics[width=0.3\textwidth]{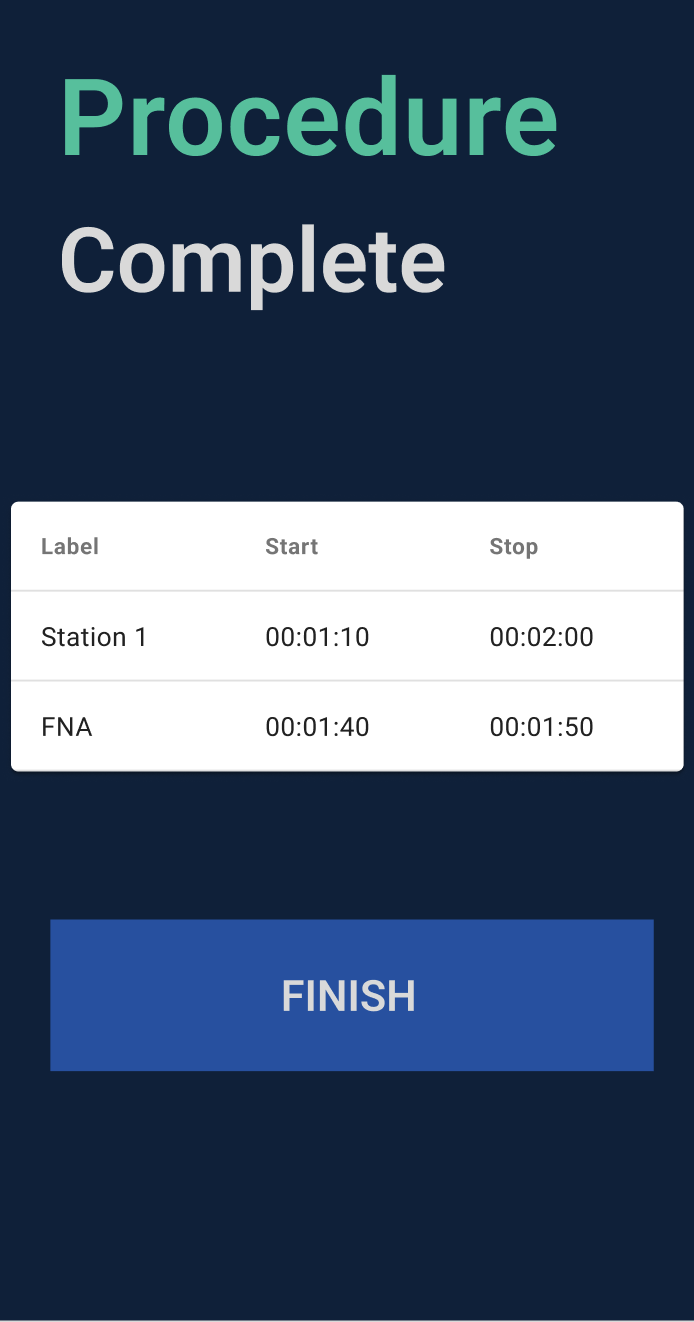}\label{fig:Label-Tool-Screen-3}}
  \captionsetup{justification=centering}
  \caption{Screenshots of the Label Tool}
  \label{fig:Label-Tool}
\end{figure}

Endoscopists from Amrita Hospitals used our open-source label tool shown in Fig~\ref{fig:Label-Tool} to annotate the timestamps during the EUS procedure. This eliminates the need for retrospective labelling. The station timestamps from the app correlate with the endoscopy machine's screen capture, allowing us to extract the corresponding frames for a station.

We used 43 clinical videos, which were captured at 24 frames per second (FPS), and their timestamps of Station 1, Station 2, and Station 3 of the Endoscopy procedure collected over three months as our dataset for this study. This dataset is cleaned and prepared for training using our data preparation step.~\ref{data-gen}. The images were extracted from EUS videos at a rate of 1FPS. Previously, we experimented with different frame rates ranging from 1 to 24, but we settled on 1FPS because other options reduced the model's performance. The dataset comprises 15,545 images, divided as 2,242 for testing and 13303 for training across all three stations, as shown in Table~\ref{tab:dataset_summary}. 

We didn't manually choose or use a special set of 'gold standard' images for training our model. The similarity between nearby interval frames means that testing may involve predicting the class of a patient already in the training set, making accurate assessment difficult. To address this, patient images were not mixed across splits. Instead, the dataset was split based on patients while maintaining balanced proportions of station images in the train, test, and validation sets. This approach evaluates the model's generalization capabilities on unseen images from different patients.

\begin{table}
\centering
\renewcommand{\arraystretch}{1.2} 
\caption{Summary of EUS Dataset}
\label{tab:dataset_summary}
\begin{tabular}{@{}crr@{}}
\toprule
\textbf{Station} & \textbf{Train} & \textbf{Test} \\ 
\midrule
Station 1 & 4,179 & 744 \\
Station 2 & 5,602 & 830 \\
Station 3 & 3,522 & 668 \\
\midrule
\textbf{Total}   & 13,303 & 2,242 \\
\bottomrule
\end{tabular}
\end{table}

\begin{figure}
  \centering
  \subfloat[Station 1]{\label{fig:figure1}\includegraphics[width=0.45\textwidth]{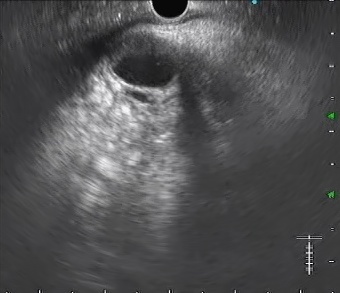}}
  \hfill
  \subfloat[Station 2]{\label{fig:figure2}\includegraphics[width=0.45\textwidth]{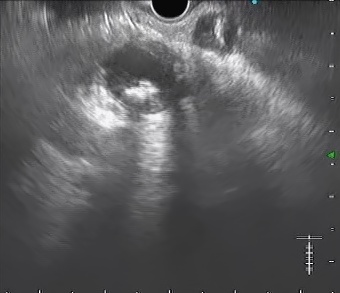}}
  \vskip\baselineskip
  \subfloat[Station 3]{\label{fig:figure3}\includegraphics[width=0.45\textwidth]{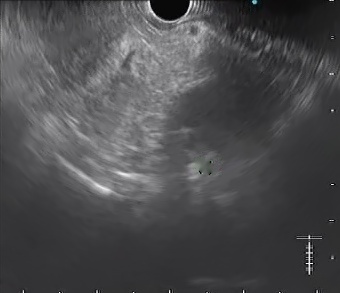}}
  \captionsetup{justification=centering}
  \caption{Images of different stations in EUS}
  \label{fig:both}
\end{figure}

\subsection{Dataset Preparation}\label{data-gen}

\begin{figure}
  \centering
  \subfloat[GUI image]{\label{fig:figure4}\includegraphics[width=0.45\textwidth]{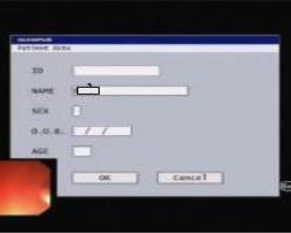}}
  \hfill
  \subfloat[white light image]{\label{fig:figure5}\includegraphics[width=0.45\textwidth]{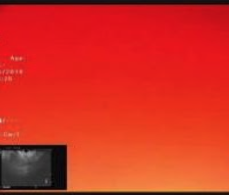}}
  \vskip\baselineskip
  \subfloat[Green pointer image]{\label{fig:figure6}\includegraphics[width=0.45\textwidth]{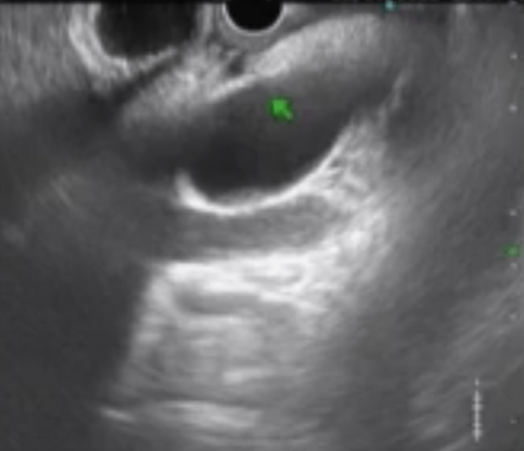}}
  \hfill
  \subfloat[Blackened image]{\label{fig:figure7}\includegraphics[width=0.45\textwidth]{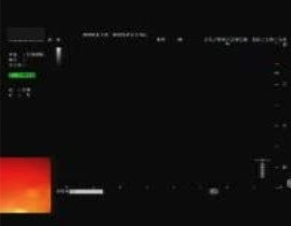}}
  \captionsetup{justification=centering}
  \caption{Different types of noises in EUS videos}
  \label{fig:four}
\end{figure}

We utilized the EUS videos of patients to generate the dataset as mentioned in the Section \ref{dataset}. We identified four distinct types of noises represented in Fig \ref{fig:four} - GUI images, white light, green pointers (used by doctors to point on the screen in real-time), and blackened images.

We performed a set of data-cleaning steps on the extracted frames. First, we removed the GUI and pink images using two histogram comparison techniques - histogram comparison intersection~\cite{histintersect} and histogram comparison Bhattacharya~\cite{bhattacharya}. Histogram comparison intersection measures the overlap between two histograms, while histogram comparison Bhattacharya measures the distance between them. These techniques are chosen because they provide a comprehensive understanding of the similarity between two histograms of two images.
It is worth noting that the specific threshold values we have used in our processes were determined through trial and error on our dataset. Thus, they may not be standard and may need to be adjusted for different datasets. 
We compared the extracted images to a reference image. If the histogram comparison intersection value is less than or equal to 1.031 and the histogram comparison Bhattacharya value is more than or equal to 0.95, we detected the pink photos. If the histogram comparison intersection value is larger than or equal to 1.42 and the histogram comparison Bhattacharya value is less than or equal to 0.18, we remove the GUI image.

Next, we removed the blackened images by computing the average intensity values of all pixels in the images. If the average pixel value is less than the threshold of 12, then that image is deemed blackened and removed. Finally, the images containing green pointers were replaced using an image processing technique called inpainting~\cite{opencv_inpainting}.

To further enhance the quality and contrast of the extracted frames, we experimented with several image enhancement algorithms, including Contrast-limited adaptive histogram equalization(CLAHE), Gaussian smoothing, denoising, quantile capping, and Fourier Transform as discussed in section \ref{prepocessing}.

\begin{figure}
\centering
\includegraphics[width=0.9\textwidth]{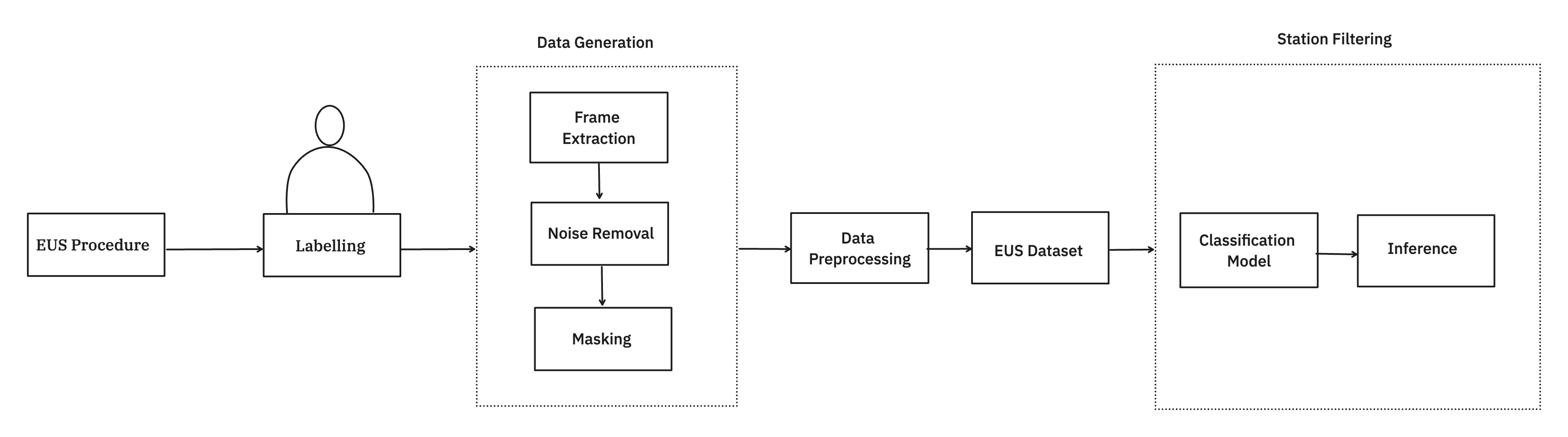}
\caption{Diagram of proposed framework system}\label{architecture}
\end{figure}

The images were normalized and standardized using the mean and standard deviation in the training set. This is to prevent any data leakage and promotes robust generalization to unseen data. The architecture diagram of the proposed framework system is displayed in Fig \ref{architecture}.

\subsection{Performance Measures}

We use balanced accuracy as the primary evaluation metric and weighted precision and weighted recall as the secondary evaluation metrics. We choose balanced accuracy as all mistakes are equally weighted; that is, all mistakes are equally important. Balanced accuracy considers the model's accuracy in each class while also considering the number of images in each class. This ensures that the performance measure is not skewed by the disproportionate number of images in each class.

\begin{equation}
\text{balanced accuracy} = \frac{1}{k}\sum_{i=1}^{k}\frac{TP_{i}}{TP_{i}+FN_{i}}
\end{equation}

where,

    $TP_{i}$: number of true positives for class $i$ 

    $FN_{i}$: number of false negatives for class $i$ 

    $k$:  total number of classes

The use of weighted precision and weighted recall allows us to account for the imbalance in the dataset by giving more weight to the classes with more images. This enables us to accurately measure the precision and recall for each class, taking into consideration the proportion of images in each class.

\begin{equation}
precision_{weighted} = \frac{n_{1}P_{1}+n_{2}P_{2}+n_{3}P_{3}}{n_{1}+n_{2}+n_{3}}
\end{equation}

\begin{equation}
recall_{weighted} = \frac{n_{1}R_{1}+n_{2}R_{2}+n_{3}R_{3}}{n_{1}+n_{2}+n_{3}}
\end{equation}

where,

    $P_{i}$: precision for class $i$

    $R_{i}$: recall for class $i$
    
    $n_{i}$: number of instances of class $i$

\section{Results and Discussion}

\begin{table}
\centering
\caption{Performance Comparisons, best Balanced Accuracy(BA) is bolded}
    \label{result-table}
\resizebox{\columnwidth}{!}{%
\begin{tabular}{@{}lrrr rrr rrr@{}}
\toprule
 &
  \multicolumn{3}{c}{Resnet18} &
  \multicolumn{3}{c}{Efficientnet\_b0} &
  \multicolumn{3}{c}{Densenet161} \\ \midrule
\multicolumn{1}{c}{Preprocessing} &
  \multicolumn{1}{c}{BA} &
  \multicolumn{1}{c}{Precision} &
  \multicolumn{1}{c}{Recall} &
  \multicolumn{1}{c}{BA} &
  \multicolumn{1}{c}{Precision} &
  \multicolumn{1}{c}{Recall} &
  \multicolumn{1}{c}{BA} &
  \multicolumn{1}{c}{Precision} &
  \multicolumn{1}{c}{Recall} \\ \midrule
\multicolumn{1}{l}{NO-PRE} &
  \multicolumn{1}{c}{85.6} &
  \multicolumn{1}{c}{85.7} &
  \multicolumn{1}{c}{85.6} &
  \multicolumn{1}{c}{82.2} &
  \multicolumn{1}{c}{82.3} &
  \multicolumn{1}{c}{82.1} &
  \multicolumn{1}{c}{\textbf{87.6}} &
  \multicolumn{1}{c}{88.5} &
  \multicolumn{1}{c}{87.7} \\
\multicolumn{1}{l}{CLAHE} &
  \multicolumn{1}{c}{\textbf{84.9}} &
  \multicolumn{1}{c}{85.8} &
  \multicolumn{1}{c}{85.3} &
  \multicolumn{1}{c}{80.3} &
  \multicolumn{1}{c}{80.8} &
  \multicolumn{1}{c}{80.5} &
  \multicolumn{1}{c}{80.7} &
  \multicolumn{1}{c}{82.3} &
  \multicolumn{1}{c}{81.0} \\
\multicolumn{1}{l}{DENOISING} &
  \multicolumn{1}{c}{83.9} &
  \multicolumn{1}{c}{84.2} &
  \multicolumn{1}{c}{84.1} &
  \multicolumn{1}{c}{80.8} &
  \multicolumn{1}{c}{81.4} &
  \multicolumn{1}{c}{81.0} &
  \multicolumn{1}{c}{\textbf{89.0}} &
  \multicolumn{1}{c}{90.0} &
  \multicolumn{1}{c}{88.9} \\
\multicolumn{1}{l}{QUANTILE CAP} &
  \multicolumn{1}{c}{75.9} &
  \multicolumn{1}{c}{76.5} &
  \multicolumn{1}{c}{76.0} &
  \multicolumn{1}{c}{\textbf{77.3}} &
  \multicolumn{1}{c}{78.0} &
  \multicolumn{1}{c}{77.5} &
  \multicolumn{1}{c}{59.8} &
  \multicolumn{1}{c}{70.0} &
  \multicolumn{1}{c}{61.6} \\
\multicolumn{1}{l}{FFT-Normal} &
  \multicolumn{1}{c}{84.4} &
 \multicolumn{1}{c}{85.8}  &
  \multicolumn{1}{c}{84.6} &
  \multicolumn{1}{c}{80.2} &
  \multicolumn{1}{c}{80.4} &
  \multicolumn{1}{c}{80.2} &
  \multicolumn{1}{c}{\textbf{88.5}} &
  \multicolumn{1}{c}{90.0} &
  \multicolumn{1}{c}{88.8} \\ 
  \multicolumn{1}{l}{GAUSSIAN Smoothing} &
  \multicolumn{1}{c}{76.2} &
  \multicolumn{1}{c}{79.7} &
  \multicolumn{1}{c}{75.6} &
  \multicolumn{1}{c}{56.3} &
  \multicolumn{1}{c}{68.2} &
  \multicolumn{1}{c}{54.5} &
  \multicolumn{1}{c}{\textbf{78.7}} &
  \multicolumn{1}{c}{84.5} &
  \multicolumn{1}{c}{79.0} \\ \bottomrule
\end{tabular}%
}
\end{table}

The results in Table \ref{result-table} summarise the outcomes of our experiments. The results of ResNet18, DenseNet161, and EfficientNetB0 trained on the datasets of preprocessing techniques are shown. DenseNet161, trained on a dataset preprocessed with Denoising, achieved the highest performance, attaining a balanced accuracy of 89\%. One surprising result is that models without preprocessing perform only 2\% lower than the best-performing model. This has a significant downstream implication, as it provides simplicity in the workflow and deployment of the model. Without any complex preprocessing steps, it will increase the inference time of the CAD system during live EUS procedures.

Furthermore, the preprocessing technique FFT outperformed the baseline by 88.53\%. Quantile capping, Gaussian smoothing, and CLAHE on the other hand, performed poorly on the dataset, with balanced accuracy of 77.39\%, 78.7\%, and 84.87\% respectively. This outcome emphasizes the importance of evaluating the applicability of preprocessing techniques on a case-by-case basis in medical imaging.

\subsection{Qualitative Analysis}
We conducted a qualitative analysis to evaluate the effectiveness of our deep-learning models in differentiating stations in the EUS procedure. We utilized the Grad-CAM technique to visualize the regions of interest (ROIs) that our model used to make predictions. Grad-CAM is a technique used in deep learning that produces visual explanations of the decision-making process of a convolutional neural network by highlighting the regions of an input image that are most important for the network's predictions.

According to our findings, the features emphasized by our deep learning model to classify the EUS images are consistent with the reference points used by our expert doctor. In essence, the model's attention mechanism appears to be focused on the same visual signals that expert doctors use in EUS recordings to diagnose anomalies.

\begin{figure}
  \centering
  \subfloat[Station-1]{\label{fig:figure18}\includegraphics[width=0.45\textwidth]{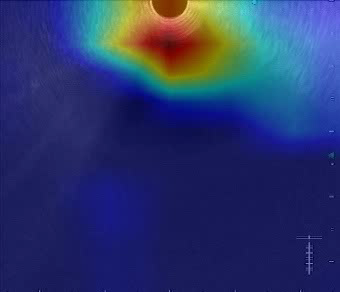}}
  \hfill
  \subfloat[Station-2]{\label{fig:figure19}\includegraphics[width=0.45\textwidth]{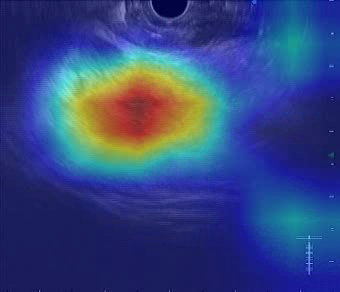}}
  \vskip\baselineskip
  \subfloat[Station-3]{\label{fig:figure20}\includegraphics[width=0.45\textwidth]{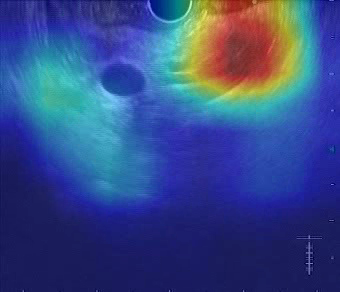}}
  \captionsetup{justification=centering}
  \caption{Grad-CAM visualisation on EUS}
  \label{fig:gradcam}
\end{figure}

\section{Conclusion}\label{conclusion}
In this work, we demonstrate that it is possible to build a CAD system to help train doctors in EUS station identification with frugal resources. We created an open-source labeling tool to annotate the timestamps of pancreas stations during the EUS procedure. This form of annotation happens in real-time during the procedure, requiring little extra effort on the behalf of the clinicians. We then demonstrate that it is possible to achieve state-of-the-art results with limited data (15\% of other studies). Notably the dataset does not have any hand annotations by doctors for gold-standard images. Instead, we train our models on the raw video footage with only the time stamps for station identification. We tested three models with a number of preprocessing techniques and found the best result was DenseNet161 with an accuracy of 89\%.
Furthermore, the result also shows that without preprocessing, the model achieved an accuracy of 87.6\%. It is desirable to drop preprocessing as it will greatly increase the speed of inference for the real-time application of the CAD.

Our results thus provide proof that a simplified approach to obtaining annotated EUS videos and using a small dataset, combined with basic deep learning techniques, can yield competitive performance. We believe in time and collaboration, we can reach much higher capacity datasets and obtain significantly improved models.

\section{Declarations}
\subsection{Data Availability}
The datasets generated during and/or analyzed during the current study are not publicly available as it is sensitive data for which we do not have consent for public distribution. Those wishing to collaborate formally are requested to email.
\subsection{Conflict of Interest}
The authors have no relevant financial or nonfinancial interests to disclose.

\bibliography{references}

\begin{thebibliography}{24}
\providecommand{\natexlab}[1]{#1}
\providecommand{\url}[1]{{#1}}
\providecommand{\urlprefix}{URL }
\providecommand{\doi}[1]{\url{https://doi.org/#1}}
\providecommand{\eprint}[2][]{\url{#2}}
 \bibcommenthead

\bibitem[{Aloysius and Geetha(2017)}]{deeplearning-review}
Aloysius N, Geetha M (2017) A review on deep convolutional neural networks. In: 2017 international conference on communication and signal processing (ICCSP), IEEE, pp 0588--0592

\bibitem[{Anwar et~al(2018)Anwar, Majid, Qayyum, Awais, Alnowami, and Khan}]{CNN}
Anwar SM, Majid M, Qayyum A, et~al (2018) Medical image analysis using convolutional neural networks: a review. Journal of medical systems 42:1--13

\bibitem[{Bray et~al(2018)Bray, Ferlay, Soerjomataram, Siegel, Torre, and Jemal}]{Bray-F-cancer-stats}
Bray F, Ferlay J, Soerjomataram I, et~al (2018) Global cancer statistics 2018: Globocan estimates of incidence and mortality worldwide for 36 cancers in 185 countries. CA: a cancer journal for clinicians 68(6):394--424

\bibitem[{Dalal and Triggs(2005)}]{bhattacharya}
Dalal N, Triggs B (2005) Histograms of oriented gradients for human detection. In: 2005 IEEE computer society conference on computer vision and pattern recognition (CVPR'05), Ieee, pp 886--893

\bibitem[{Faulx et~al(2017)Faulx, Lightdale, Acosta, Agrawal, Bruining, Chandrasekhara, Eloubeidi, Gurudu, Kelsey, Khashab et~al}]{faulx2017guidelines}
Faulx AL, Lightdale JR, Acosta RD, et~al (2017) Guidelines for privileging, credentialing, and proctoring to perform gi endoscopy. Gastrointestinal endoscopy 85(2):273--281

\bibitem[{Fleurentin et~al(2022)Fleurentin, Mazellier, Meyer, Montanelli, Swanstrom, Gallix, Sosa~Valencia, and Padoy}]{automatic-video}
Fleurentin A, Mazellier JP, Meyer A, et~al (2022) Automatic pancreas anatomical part detection in endoscopic ultrasound videos. Computer Methods in Biomechanics and Biomedical Engineering: Imaging \& Visualization pp 1--7

\bibitem[{Gonzalo-Marin et~al(2014)Gonzalo-Marin, Vila, and Perez-Miranda}]{Gonzalo-Marin-Role-endoscopic}
Gonzalo-Marin J, Vila JJ, Perez-Miranda M (2014) Role of endoscopic ultrasound in the diagnosis of pancreatic cancer. World journal of gastrointestinal oncology 6(9):360

\bibitem[{Heo et~al(2020)Heo, Kim, and Lee}]{heo2020image}
Heo YC, Kim K, Lee Y (2020) Image denoising using non-local means (nlm) approach in magnetic resonance (mr) imaging: a systematic review. Applied Sciences 10(20):7028

\bibitem[{Jaramillo et~al(2022)Jaramillo, Ruano, G{\'o}mez, and Romero}]{María-Jaramillo}
Jaramillo M, Ruano J, G{\'o}mez M, et~al (2022) Automatic detection of pancreatic tumors in endoscopic ultrasound videos using deep learning techniques. In: Medical Imaging 2022: Ultrasonic Imaging and Tomography, SPIE, pp 106--115

\bibitem[{Keerthi and Santhi(2023)}]{precise-image-preprocessing}
Keerthi S, Santhi P (2023) Precise multi-class classification of brain tumor via optimization based relevance vector machine. Intelligent Automation \& Soft Computing 36(1)

\bibitem[{Lee et~al(2005)Lee, Xin, and Westland}]{histintersect}
Lee S, Xin JH, Westland S (2005) Evaluation of image similarity by histogram intersection. Color Research \& Application: Endorsed by Inter-Society Color Council, The Colour Group (Great Britain), Canadian Society for Color, Color Science Association of Japan, Dutch Society for the Study of Color, The Swedish Colour Centre Foundation, Colour Society of Australia, Centre Fran{\c{c}}ais de la Couleur 30(4):265--274

\bibitem[{Liu et~al(2019)Liu, Faes, Kale, Wagner, Fu, Bruynseels, Mahendiran, Moraes, Shamdas, Kern et~al}]{comparison-of-deep-learning}
Liu X, Faes L, Kale AU, et~al (2019) A comparison of deep learning performance against health-care professionals in detecting diseases from medical imaging: a systematic review and meta-analysis. The lancet digital health 1(6):e271--e297

\bibitem[{LU et~al(2021)LU, WU, YAO, CHEN, and YU}]{Zhang-J-segmentation}
LU Z, WU H, YAO L, et~al (2021) A station recognition and pancreatic segmentation system in endoscopic ultrasonography based on deep learning. Chinese Journal of Digestive Endoscopy pp 778--782

\bibitem[{Mohan et~al(2020)Mohan, Nair et~al}]{fuzzy-clahe}
Mohan M, Nair LS, et~al (2020) Fuzzy c-means segmentation on enhanced mammograms using clahe and fourth order complex diffusion. In: 2020 Fourth International Conference on Computing Methodologies and Communication (ICCMC), IEEE, pp 647--651

\bibitem[{{OpenCV}(Year)}]{opencv_inpainting}
{OpenCV} (Year) Inpainting tutorial - opencv documentation. Online, \urlprefix\url{https://docs.opencv.org/3.4/df/d3d/tutorial_py_inpainting.html}

\bibitem[{Osareh and Shadgar(2011)}]{CAD-2}
Osareh A, Shadgar B (2011) A computer aided diagnosis system for breast cancer. International Journal of Computer Science Issues (IJCSI) 8(2):233

\bibitem[{Panwar et~al(2020)Panwar, Gupta, Siddiqui, Morales-Menendez, Bhardwaj, and Singh}]{gradcam-for-medical}
Panwar H, Gupta P, Siddiqui MK, et~al (2020) A deep learning and grad-cam based color visualization approach for fast detection of covid-19 cases using chest x-ray and ct-scan images. Chaos, Solitons \& Fractals 140:110,190

\bibitem[{Pizer(1990)}]{Pizer-clahe}
Pizer SM (1990) Contrast-limited adaptive histogram equalization: Speed and effectiveness stephen m. pizer, r. eugene johnston, james p. ericksen, bonnie c. yankaskas, keith e. muller medical image display research group. In: Proceedings of the first conference on visualization in biomedical computing, Atlanta, Georgia, p~1

\bibitem[{Rawla et~al(2019)Rawla, Sunkara, and Gaduputi}]{Rawla-P-Epidemiology}
Rawla P, Sunkara T, Gaduputi V (2019) Epidemiology of pancreatic cancer: global trends, etiology and risk factors. World journal of oncology 10(1):10--27

\bibitem[{Selvaraju et~al(2017)Selvaraju, Cogswell, Das, Vedantam, Parikh, and Batra}]{Grad-CAM}
Selvaraju RR, Cogswell M, Das A, et~al (2017) Grad-cam: Visual explanations from deep networks via gradient-based localization. In: Proceedings of the IEEE international conference on computer vision, pp 618--626

\bibitem[{Tan and Le(2019)}]{EfficientNet}
Tan M, Le Q (2019) Efficientnet: Rethinking model scaling for convolutional neural networks. In: International conference on machine learning, PMLR, pp 6105--6114

\bibitem[{Xiao et~al(2010)Xiao, Ho, and Bargiela}]{Xiao-gaussian}
Xiao K, Ho SH, Bargiela A (2010) Automatic brain mri segmentation scheme based on feature weighting factors selection on fuzzy c-means clustering algorithms with gaussian smoothing. International Journal of Computational Intelligence in Bioinformatics and Systems Biology 1(3):316--331

\bibitem[{Yao et~al(2021)Yao, Zhang, Liu, Zhu, Ding, Chen, Wu, Lu, Zhou, Zhang et~al}]{bile-duct}
Yao L, Zhang J, Liu J, et~al (2021) A deep learning-based system for bile duct annotation and station recognition in linear endoscopic ultrasound. EBioMedicine 65:103,238

\bibitem[{Zhang et~al(2020)Zhang, Zhu, Yao, Ding, Chen, Wu, Lu, Zhou, Zhang, An et~al}]{Zhang-Jun-video}
Zhang J, Zhu L, Yao L, et~al (2020) Deep learning--based pancreas segmentation and station recognition system in eus: Development and validation of a useful training tool (with video). Gastrointestinal endoscopy 92(4):874--885

\end{thebibliography}


\end{document}